\documentclass[12pt]{article}
\pdfoutput=1
\usepackage{amssymb,amsmath}
\usepackage{graphicx}
\usepackage{color}
\usepackage{verbatim}

\usepackage[colorlinks=true
,urlcolor=blue
,citecolor=blue
,linkcolor=blue
,pagecolor=blue
,linktocpage=true
,pdfproducer=medialab
]{hyperref}
\usepackage[a4paper,width=17.1cm]{geometry}
\makeatletter \renewcommand{\@dotsep}{10000} \makeatother
\usepackage{appendix}
\usepackage{subcaption}
\usepackage{mwe}

\newcommand{\gmu}{\ensuremath{(g-2)_{\mu}}}
\newcommand{\damu}{\ensuremath{\Delta a_{\mu}}}

\begin{document}

\begin{center}

 {\Large  \textbf{Muon g-2 and Dark Matter\\ \vspace{2mm} in the NUGM + NUHM2 model} 
 } \vspace{1cm}

{  M. Adeel Ajaib$^{a,}$\footnote{ E-mail: adeel@udel.edu}
 , Fariha Nasir$^{a,}$\footnote{ E-mail: fariha@udel.edu}} \vspace{.5cm}

{ \it 
$^a$Pennsylvania State University, Abington PA, United States
 \\
} \vspace{.5cm}

\vspace{1.5cm}
 {\bf Abstract}\end{center}

We analyze the NUGM + NUHM2 model and study the implications of recent experimental constraints on the parameter space of this model. We also explore the nature of dark matter in this model and present the parameter space characterized by various compositions of the neutralino. The region of the parameter space that satisfies the observed deviation in the anomalous magnetic moment of the muon is explored more rigorously. We also present results corresponding to the production cross section and decay widths of a light CP-even Higgs boson and observe that there is a narrow region of the parameter space corresponding to a light stau which explains the observed deviation in the muon g-2 and also leads to enhancement in the Higgs production and decay widths.

\newpage

\renewcommand{\thefootnote}{\arabic{footnote}}
\setcounter{footnote}{0}

\section{Introduction}

The lack of experimental evidence continues to constrain the parameter space of Supersymmetry (SUSY). There are however several reasons why SUSY is still a viable candidate for physics beyond the Standard Model (SM) as it can be employed to solve some of the issues in the Standard Model (SM) such as the gauge hierarchy problem and the unification of gauge couplings. Moreover, the Minimal Supersymmetric Standard Model (MSSM) with R-parity entails a cold dark matter candidate and a 125 GeV Higgs boson can be readily accommodated in various SUSY models. The ATLAS and CMS experiments have performed various searches for SUSY particles but, to date, no signals have been detected by these experiments. Run 2 of the Large Hadron Collider (LHC) with an integrated luminosity of $\sim$ 139 fb$^{-1}$ require a limit of $m_{\tilde{g}}  \gtrsim  2.3~{\rm TeV}$ on the gluino masses and the top squark masses limits are  $m_{\tilde{t}_1}  \gtrsim  1.2 ~{\rm TeV}$, respectively. LHC Run 3 started in July 2022 delivering proton-proton collisions with center-of-mass energies of 13.6 TeV and the search for SUSY signals continues. Several recent studies have considered the affect of these constraints on the parameter space of various SUSY models \cite{Ellis:2022emx}.

In addition to the experiments at the LHC, Fermilab's Muon experiment is also at the forefront of testing physics beyond the SM. Its measurement of the anomalous magnetic moment of the muon can also be an indication of new physics beyond the SM. The observed $4.2 \sigma$ deviation in the muon anomalous magnetic moment
$a_{\mu}=(g-2)_{\mu}/2$ (muon $g-2$) from its SM prediction \cite{Muong-2:2021ojo}
\begin{eqnarray}
\label{gg-22}
\Delta a_{\mu}\equiv a_{\mu}({\rm exp})-a_{\mu}({\rm SM})= (25.1 \pm 5.9) \times 10^{-10}.
\end{eqnarray}
There have been several studies in recent year that have attempted to explain this long standing anomaly  \cite{Wang:2021bcx}. In this paper we revisit the Non-Universal Gaugino Mass and the Non-Universal Higgs mass (NUGM + NUHM2)  models and analyze the implications of constraints from direct and indirect experiments. In addition, we will also explore the parameter space which explains the observed deviation in $\gmu$ given in equation (1). We will explore the nature of the neutralino in the model's parameter space and present results for different coannihilation scenarios. Furthermore, we will also present results for the Higgs production cross section and decay widths. We will highlight the region of the parameter space that satisfies the $\gmu$ constraint and also predicts an enhancement in the diphoton channel.

The paper is orgranized as follows: In section \ref{g-2}, we briefly review the SUSY contribution to the muon anomalous magnetic moment and present the expression for $\damu$. In section \ref{sec:parameter} we describe the scanning procedure, the constraints we implement and the parameter space of the models we study. In section \ref{sec:results}, we display and discuss the results of our parameter space scan. We conclude in section \ref{sec:conclude}

\section{\label{g-2}The Muon Anomalous Magnetic Moment}

In the MSSM, the dominant contribution to the muon anomalous magnetic moment arises from neutralino-charge slepton and chargino-sneutrino loop diagrams given by \cite{Moroi:1995yh, Martin:2001st}:

\begin{eqnarray}
\label{eqq1}
\Delta a_\mu &=& \frac{\alpha \, m^2_\mu \, \mu\,  \tan\beta}{4\pi} {\bigg \{ }
\frac{M_{2}}{ \sin^2\theta_W \, m_{\tilde{\mu}_{L}}^2}
\left[ \frac{f_{\chi}(M_{2}^2/m_{\tilde{\mu}_{L}}^2)-f_{\chi}(\mu^2/m_{\tilde{\mu}_{L}}^2)}{M_2^2-\mu^2} \right] 
\nonumber\\
&+&
\frac{M_{1} }{ \cos^2\theta_W \, (m_{\tilde{\mu}_{R}}^2 - m_{\tilde{\mu}_{L}}^2)}
\left[\frac{f_{N}(M^2_1/m_{\tilde{\mu}_{R}}^2)}{m_{\tilde{\mu}_{R}}^2} - \frac{f_{N}(M^2_1/m_{\tilde{\mu}_{L}}^2)}{m_{\tilde{\mu}_{L}}^2}\right] \, {\bigg \} },
\end{eqnarray}
 where $\alpha$ is the fine-structure constant, $m_\mu$ is the muon mass, $\mu$ denotes  the bilinear Higgs mixing term, and $\tan\beta$ is the ratio of the vacuum expectation values (VEV) of the MSSM Higgs doublets. $M_1$ and $M_2$ denote the $U(1)_Y$ and $SU(2)$ gaugino masses respectively, $\theta_W$  is the weak mixing angle, and $m_{\tilde{\mu}_{L}}$ and $m_{\tilde{\mu}_{R}}$ are the left and right handed smuon masses. The loop functions are defined as follows:
\begin{eqnarray}
f_{\chi}(x) &=& \frac{x^2 - 4x + 3 + 2\ln x}{(1-x)^3}~,\qquad ~f_{\chi}(1)=-2/3, \\
f_{N}(x) &=& \frac{ x^2 -1- 2x\ln x}{(1-x)^3}\,,\qquad\qquad f_{N}(1) = -1/3 \, .
\label{eqq2}
\end{eqnarray}

The first term in equation (\ref{eqq1}) stands for the dominant contribution coming from one loop diagram with charginos (Higgsinos and Winos), while the second term describes inputs from the bino-smuon loop.

\section{Non-Universal Gaugino and Higgs Masses}\label{sec:parameter}

It has been shown that non-universal gaugino masses  can be generated from an F-term which
is a linear combination of two distinct fields of different dimensions. \cite{Martin:2013aha}.  Another possibility is to consider two distinct sources for supersymmetry breaking \cite{Anandakrishnan:2013cwa}. With many distinct possibilities available for realizing non-universal gaugino masses at $M_{\rm GUT}$, we employ three independent masses for the  MSSM  gauginos in our study. The NUGM model is an extensively studied model and there have been several recent studies exploring this model \cite{Ajaib:2017zba}. In the Non-Universal Higgs Model 2 (NUHM2) \cite{Baer:2021aax}, the universality of the scalar masses is relaxed compare to the CMSSM and the Higgs Soft SUSY-Breaking (SSB) mass terms are assumed to be independent parameters at the GUT scale ($m_{H_u}^2 \neq m_{H_d}^2$). We choose the following ranges for the parameters of this model:
\begin{eqnarray}
 0\leq & m_{0} & \leq 30\, \textrm{TeV} \nonumber \\
 -3 \le & A_0/m_{0} & \le 3 \nonumber \nonumber \\
 2 \le & \tan\beta &  \le 60 \nonumber \\
 0 \le & M_1 &\le {\mathrm {5 \ TeV} } \nonumber \\
0 \le & M_2 &\le {\mathrm {5 \ TeV} } \nonumber \\
{\mathrm {-5 \ TeV} } \le & M_3 &\le 0  \nonumber  \\
0 \le & m_{H_u} &\le {\mathrm {5 \ TeV} } \nonumber  \\
0 \le &  m_{H_d} &\le {\mathrm {5 \ TeV} } \nonumber  \\
 &\mu >0 \nonumber 
\end{eqnarray}
Here  $M_{1}$, $M_{2}$, and $M_{3}$ denote the SSB gaugino masses for $U(1)_{Y}$, $SU(2)_{L}$ and $SU(3)_{c}$ respectively. $ \tan\beta $ is the ratio of the vacuum expectation values (VEVs) of the two MSSM Higgs doublets, and $ A_{0} $ is the universal SSB trilinear scalar interaction (with corresponding Yukawa coupling factored out).   In order to obtain the correct sign for the desired contribution to $ \gmu $, we set same signs for the parameters $\mu$, $M_1$ and $M_2$. For the top quark mass, we use {the} value $m_t = 173.3\, {\rm GeV}$.

\renewcommand{\arraystretch}{1.5}
\begin{table}[t!]
\begin{center}
\begin{tabular}{|p{11cm}|p{3cm}|}

\hline 
       Constraint          	&	 Ref.	 		 	\\
\hline

$123~{\rm GeV} \leq  m_h \leq 127~{\rm GeV}$       	& \cite{:2012gk,:2012gu}\\
\hline							

$ 0.8 \times 10^{-9} \leq BR(B_s \rightarrow \mu^+ \mu^-)  \leq\, 6.2 \times 10^{-9} \;
 (2\sigma)$       	& \cite{Aaij:2012nna}\\
\hline							
$2.99 \times 10^{-4} \leq BR(b \rightarrow s \gamma)  \leq\, 3.87 \times 10^{-4} \;
 (2\sigma)$      	& \cite{Amhis:2012bh}\\
\hline							
$0.15 \leq \frac{BR(B_u\rightarrow
\tau \nu_{\tau})_{\rm MSSM}}{BR(B_u\rightarrow \tau \nu_{\tau})_{\rm SM}} \leq\, 2.41 \;
(3\sigma)$       	&  \cite{Asner:2010qj}\\
\hline							
$m_{\tilde{g}}  \gtrsim  2.3~{\rm TeV}$, $m_{\tilde{q}}  \gtrsim  2 ~{\rm TeV}$     	&\cite{Aaboud:2017vwy,Vami:2019slp,Sirunyan:2017kqq}\\
\hline							
$m_{\tilde{\chi}^\pm}  \gtrsim  430~{\rm GeV}$      	& \cite{lhc-chargino}.\\
\hline

\hline  
\end{tabular}
\end{center}
\caption{Table of constraints we apply in our analysis }
\label{table0}
\end{table}
\renewcommand{\arraystretch}{1}

{We employ Isajet~7.84 \cite{ISAJET} interfaced with Micromegas 2.4 \cite{Belanger:2008sj} and FeynHiggs \cite{Bahl:2018qog} to perform random scans over the parameter space of this model.} We use Micromegas to calculate the relic density and $BR(b \rightarrow s \gamma)$. The Higgs cross section and decay rates have been calculated using FeynHiggs.  Further details regarding our scanning procedure can be found in \cite{Ajaib:2015ika}.  After collecting the data, we impose the experimental constraints given in Table \ref{table0} on the parameter space.
In addition we subsequently implement the following two constraints on the parameter space that satisfies the above constraints:

\begin{equation}
   20 \times 10^{-10} <  \Delta a_\mu  < 30 \times 10^{-10} \,\,  (1\sigma)
\end{equation}

\begin{equation}   
  \Omega h^2< 1   \
\end{equation}

The parameter space consistent with the above $\Delta a_\mu$ constraint is scanned more rigorously. We consider a wider region of the parameter space because of the uncertainties involved in the calculations of the relic density in different spectrum codes. The benchmark scenarios we present are going to satisfy the WMAP relic density constraint given by  \cite{Akrami:2018vks}:

\begin{equation}
 0.114 < \Omega h^2 < 0.116
\end{equation}


\begin{figure}
        \includegraphics[width=8cm, height=6.5cm]{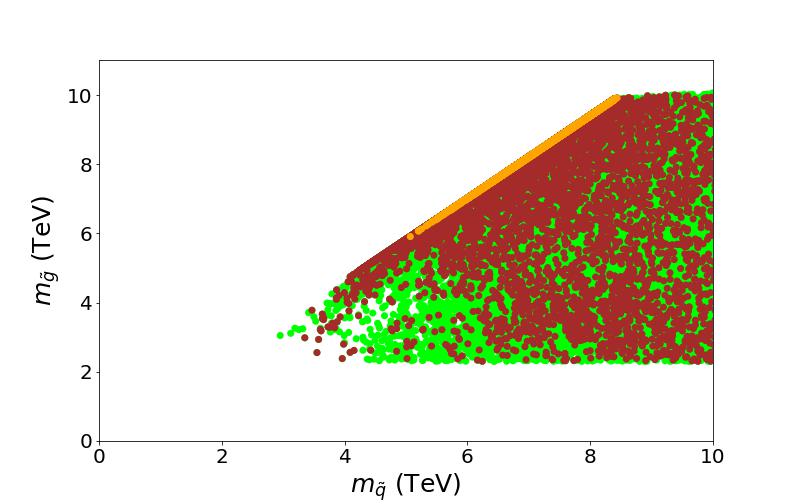}\hfill
        \includegraphics[width=8cm, height=6.5cm]{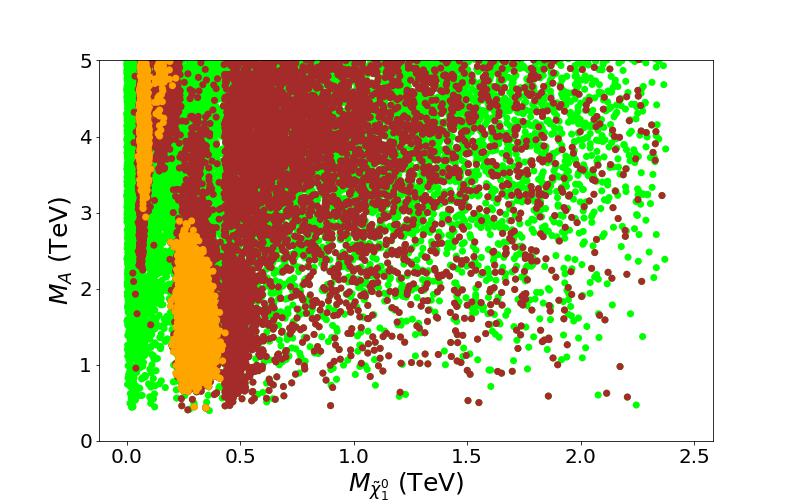}\vspace{2mm} \\ 
        \includegraphics[width=8cm, height=6.5cm]{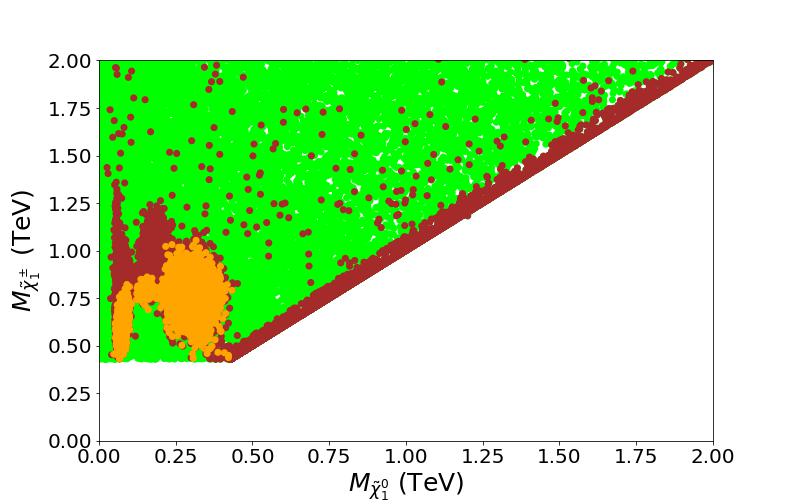}\hfill
        \includegraphics[width=8cm, height=6.5cm]{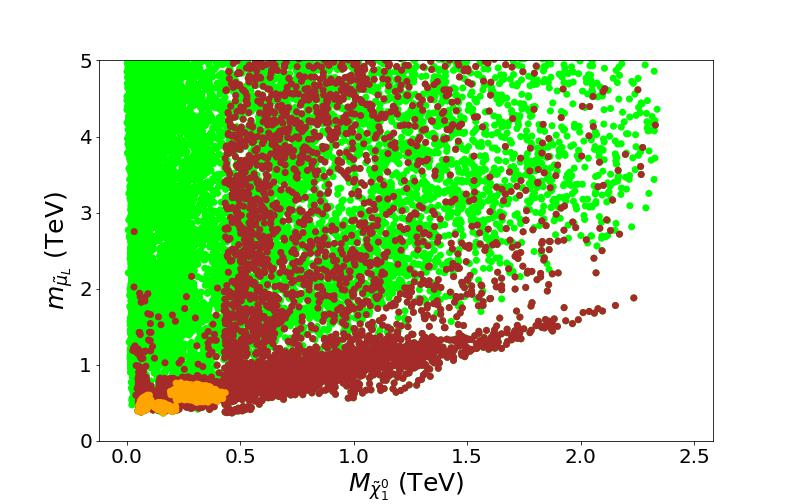}\vspace{2mm} \\ 
         \includegraphics[width=8cm, height=6.5cm]{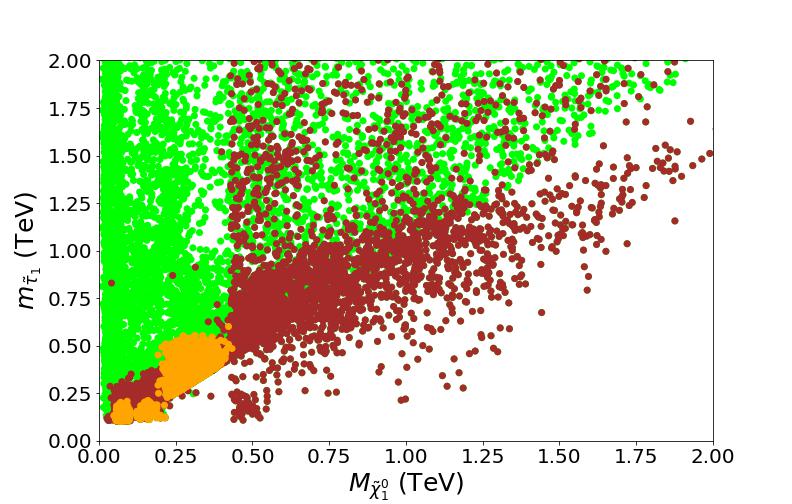}\hfill
         \includegraphics[width=8cm, height=6.5cm]{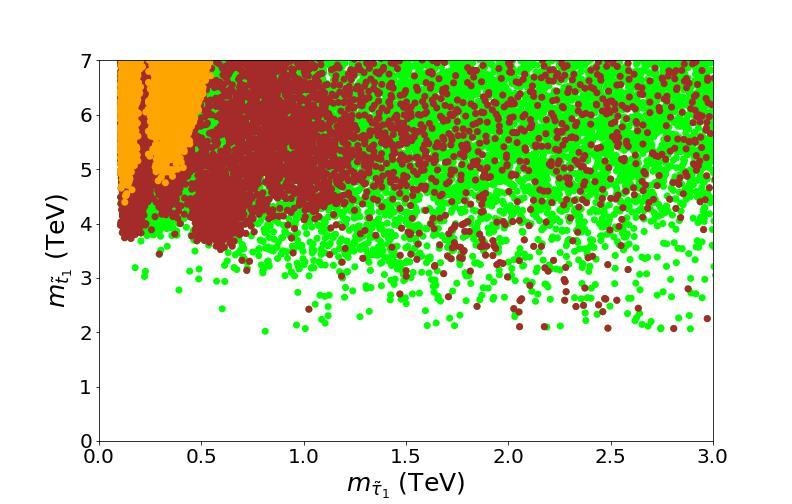}
\caption{Plots in the $m_{\tilde{g}}$ vs. $m_{\tilde{q}}$, $M_A$ vs. $M_{\tilde{\chi}_1^0}$, $M_{\tilde{\chi}_1^\pm}$ vs. $m_{\tilde{\chi}_1^0}$, $m_{\tilde{\mu}_L}$ vs. $M_{\tilde{\chi}_1^0}$, $m_{\tilde{\mu}_L}$ vs. $M_{\tilde{\chi}_1^0}$ and $m_{\tilde{t}_1}$ vs. $m_{\tilde{\tau}_1}$ planes.  \textit{Green} points satisfy the sparticle mass constraints and B-physics constraints described in Section \ref{sec:parameter}. \textit{Brown} points form a subset of the \textit{green} points and satisfy $0.001 \le \Omega h^2 \le 1 $. \textit{Orange} points are subset of the green points and satisfy the muon $g-2$ constraint described in Section  \ref{sec:parameter}.}
\label{fig:sparticle}
\end{figure}

\section{Results and Analysis}\label{sec:results}

In this section, we present our results for the parameter space scan given in the previous section. In Figure \ref{fig:sparticle}, shows our results in the $m_{\tilde{g}}$ vs. $m_{\tilde{q}}$, $M_A$ vs. $M_{\tilde{\chi}_1^0}$, $M_{\tilde{\chi}_1^\pm}$ vs. $m_{\tilde{\chi}_1^0}$,  $m_{\tilde{\mu}_L}$ vs. $M_{\tilde{\chi}_1^0}$, $m_{\tilde{\tau}_1}$ vs. $M_{\tilde{\chi}_1^0}$ and $m_{\tilde{t}_1}$ vs. $m_{\tilde{\tau}_1}$ planes. We can see that the $\gmu$ constraint implies heavy colored sparticle masses bounded in narrow intervals, namely, 
$5 {\rm \ TeV} \lesssim m_{\tilde{q}} \lesssim 8 {\rm \ TeV}$,
$6 {\rm \ TeV} \lesssim m_{\tilde{g}} \lesssim 10 {\rm \ TeV}$,
$ m_{\tilde{t}_1} \gtrsim 4.5 {\rm \ TeV}$. We use the following convention for all the following plots unless stated otherwise:

\begin{figure}[t!]
\begin{center}
        \includegraphics[width=11cm, height=7.5cm]{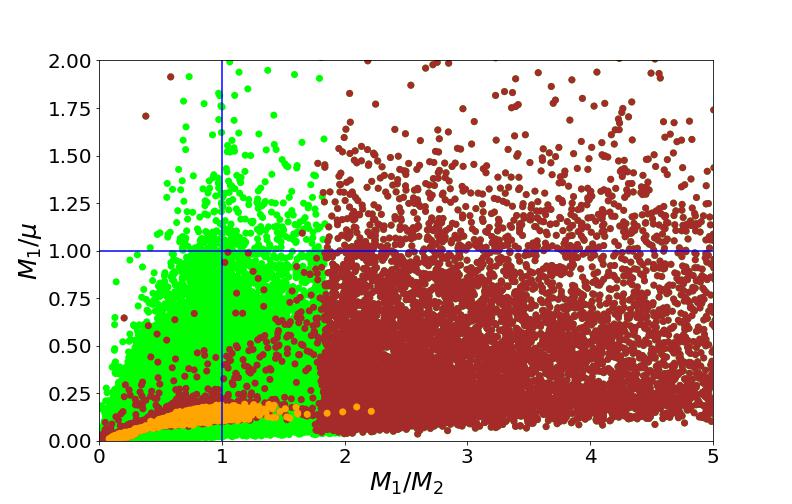}
\end{center}

\caption{Plots in the $M_1/\mu$ vs.  $M_1/M_2$ planes. Color coding is the same as in Figure \ref{fig:sparticle}.}
\label{fig:m1mu-m1m2}
\end{figure}
\begin{itemize}
\item \textit{Green} points satisfy the sparticle mass constraints and B-physics constraints described in Section \ref{sec:parameter}

\item \textit{Brown} points form a subset of the \textit{green} points and satisfy $0.001 \le \Omega h^2 \le 1 $

\item \textit{Orange} points are a subset of the \textit{green} points and satisfy the muon $g-2$ constraint described in Section  \ref{sec:parameter}.

\end{itemize}


\begin{figure}
        \includegraphics[width=8cm, height=6.5cm]{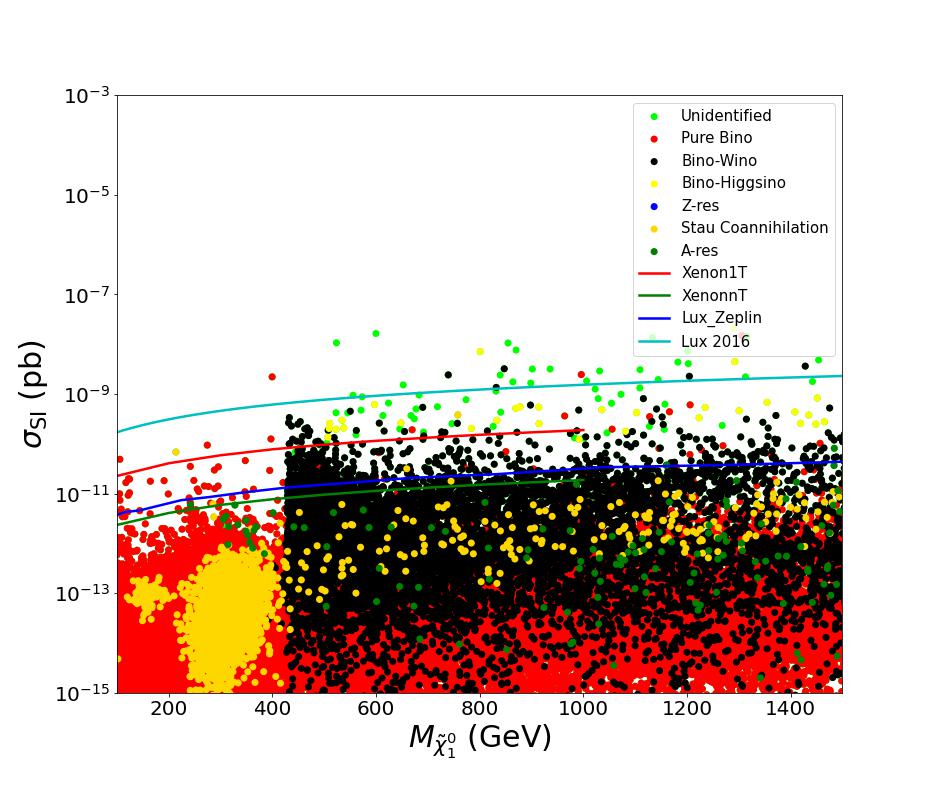}\hfill
        \includegraphics[width=8cm, height=6.5cm]{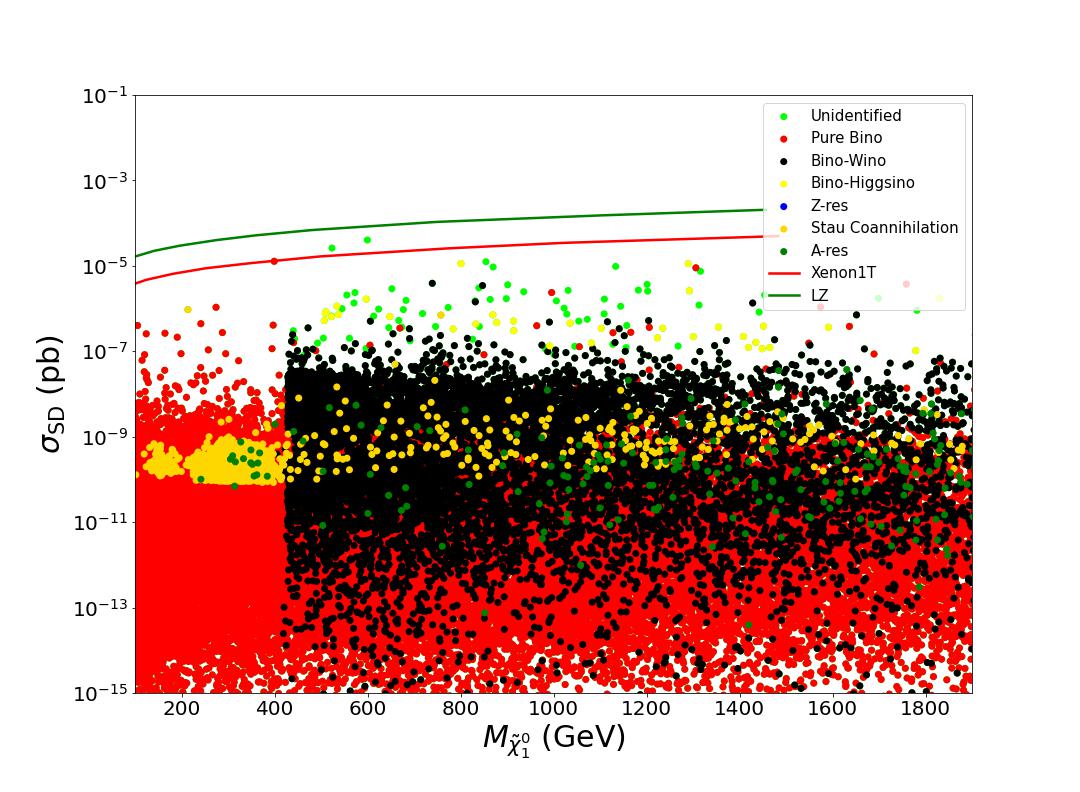} \\
        \includegraphics[width=8cm, height=6.5cm]{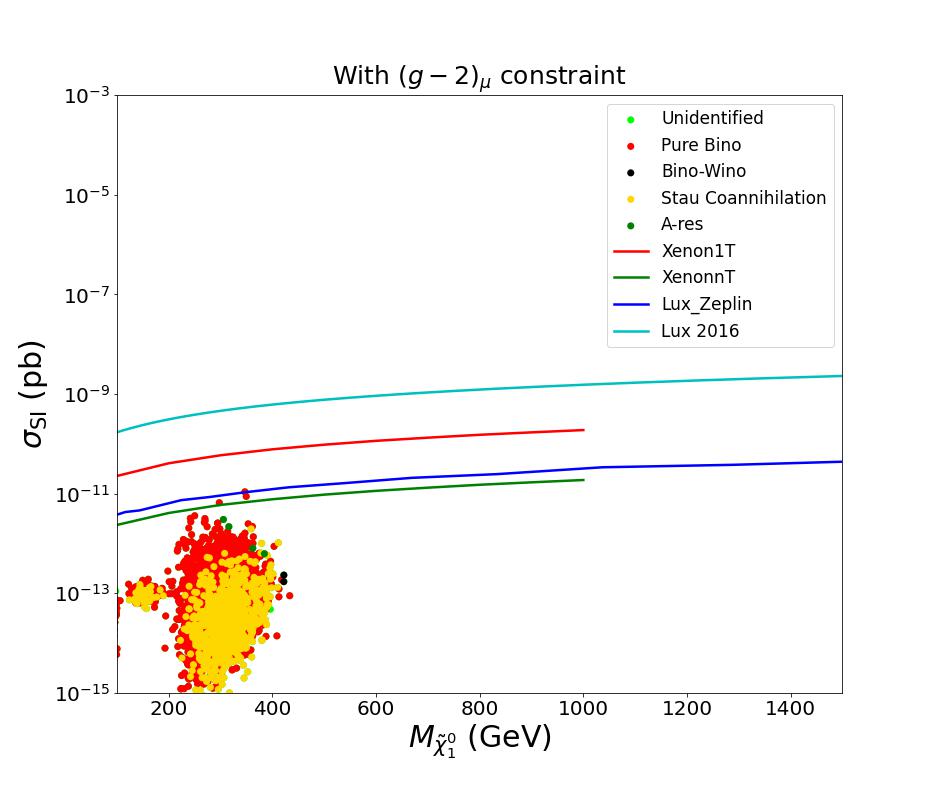}\hfill
        \includegraphics[width=8cm, height=6.5cm]{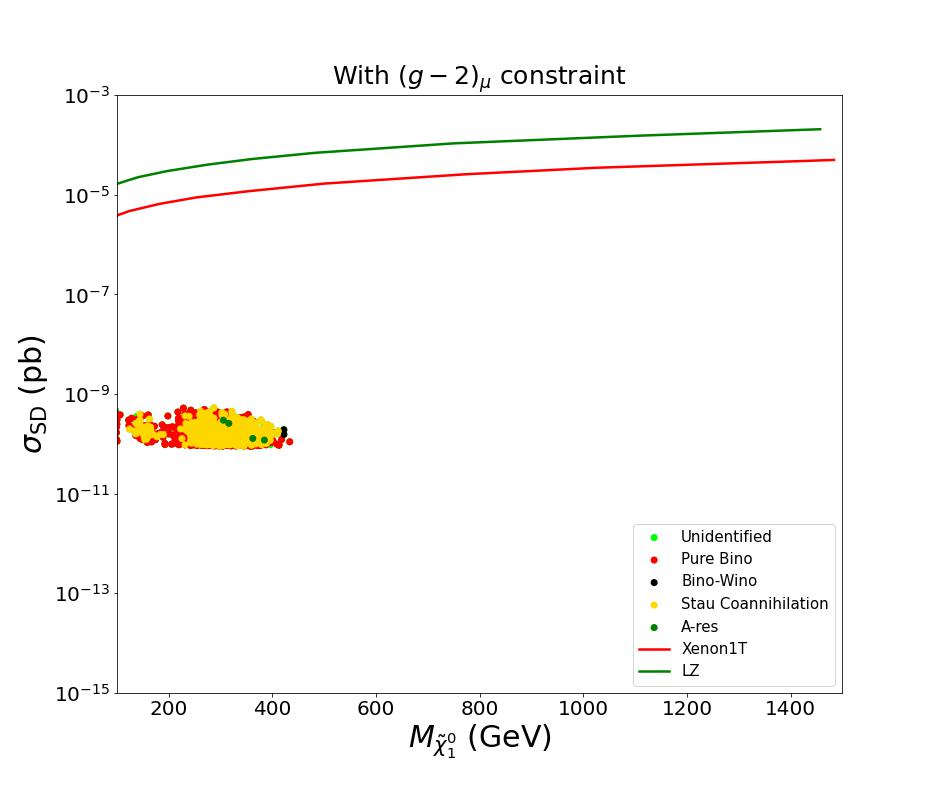} \\ 
\caption{Plots in the  $\sigma_{SI}$ vs. $M_{\tilde{\chi}_1^0}$ and $\sigma_{SD}$ vs. $M_{\tilde{\chi}_1^0}$ planes. Color coding represents the composition of the neutralino and different coannihilation channels (as shown in the plot legend).}
\label{fig:sisd}
\end{figure}

LHC Run 3 started in July 2022 delivering proton-proton collisions with center-of-mass energies of 13.6 TeV.
A portion of the parameter space of this model may be accessible at LHC Run 3 energies. We can see from the  $M_A$ vs. $M_{\tilde{\chi}_1^0}$ that the pseudoscalar Higgs boson mass can be as light 500 GeV, which is within the reach of the LHC. In the lower right panel, we can see that the g-2 constraint predicts a light stau and its mass is bounded in the interval $100 {\rm \ GeV} \lesssim m_{\tilde{\tau}_1} \lesssim 600 {\rm \ GeV}$.

Figure \ref{fig:m1mu-m1m2} gives an overview of the different types of neutralino compositions in the parameter space we study. In the region $M_1/\mu < 1$ and $M_1/M_2 < 1$, the neutralino is expected to essentially be a pure bino. It is well known in literature that a pure bino yields a large relic abundance due to small cross sections involved. However, coannihilation of the bino with other sparticles can yield the correct relic abundance particularly if the neutralino acquires a wino or a higgsino component \cite{Ajaib:2015yma}. This is possible in our analysis since the region $M_1/\mu < 1$ and $M_1/M_2 > 1 $ corresponds to a mixed bino-wino type neutralino. In the parameter space we explore, the neutralino is predominantly a bino or a mixed bino-wino type. 


\begin{figure}
        \includegraphics[width=9cm, height=6.5cm]{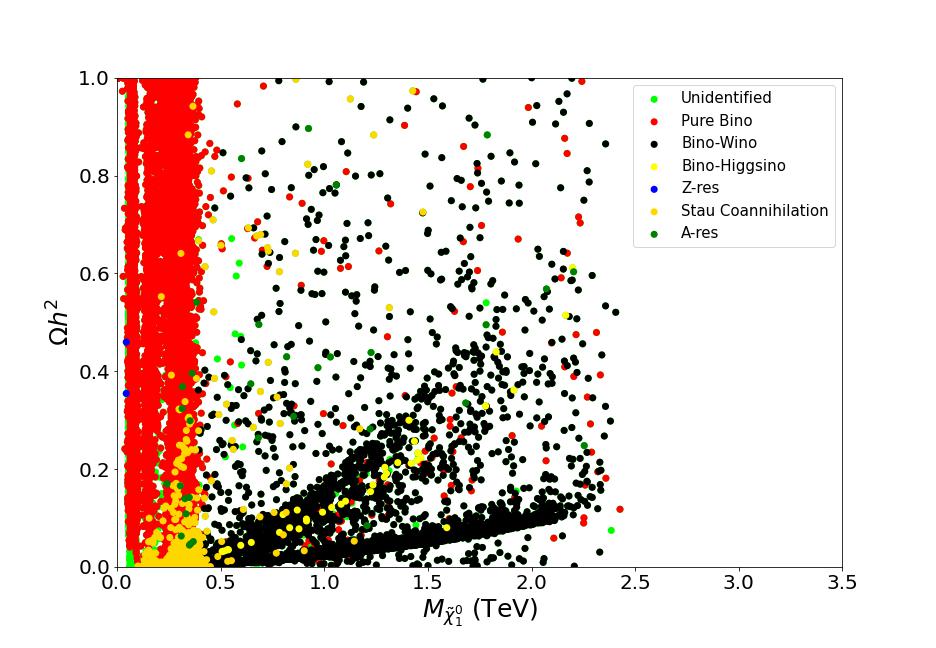}\hfill
        \includegraphics[width=9cm, height=6.5cm]{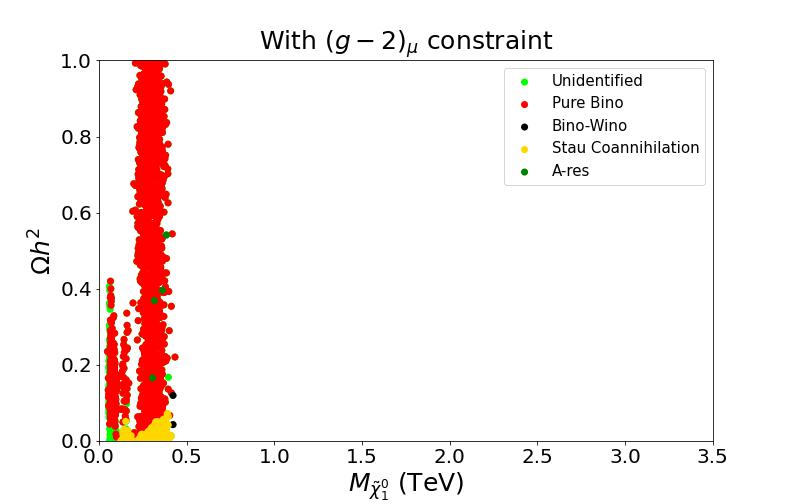} \\ 
\caption{Plots in the $\Omega h^2$ vs. $M_{\tilde{\chi}_1^0}$ planes. The color coding corresponds to the various solutions of LSP neutralino to be a dark matter candidate.
}
\label{fig:oh2}
\end{figure}

In Figure \ref{fig:oh2}, we present the parameter space in the  $\Omega h^2$ vs. $M_{\tilde{\chi}_1^0}$ plane. The color coding in the Figures corresponds to the various composition of the LSP neutralino. We can see from the left panel that there are several co-annihilation channels such as the chargino-neutralino, smuon-neutralino and stau-neutralino coannihilation channels that come into play in order to yield the desired relic abundance. We can see that the neutralino is predominantly a bino (red points) and a notable portion of the parameter space corresponds the neutralino being a mixed bino-wino type. All the points in the right panel satisfy the $\gmu$ constraint. The points corresponding to stau-coannihilation (yellow points) lead to a small relic density ($\Omega h^2 \lesssim 0.1$) for dark matter. 

In Figure \ref{fig:sisd}, we present the prospects of direct detection of neutralino dark matter in the $\sigma_{SI}$ vs. $M_{\tilde{\chi}_1^0}$ plane (left panels). The top panel displays the neutralino decomposition along with the different co-annihilation channels. The bottom left panel shows the parameter space that satisfies the $\gmu$ constraint. The solid lines represent the XENON1T~\cite{Aprile:2015uzo}, XENONnT~\cite{Aprile:2015uzo}, Lux-Zeplin ~\cite{Akerib:2015cja} and LUX2016 bound~\cite{Akerib:2016vxi}. We can observe that the parameter space of this model, particularly the region corresponding to a bino-wino type neutralino, can be tested by direct detection experiments. A portion of parameter space is already excluded by the Xenon and LUX experiments. A notable region of the parameter space with considerably low cross sections is not accessible to any of the projected sensitivities of these experiments. The parameter space consistent with the $\gmu$ constraint (lower left panel) corresponds to low cross sections for a bino and will be accessible to these experiments in the near future. 

The spin-dependent neutralino cross section is displayed in the right panels of Figure \ref{fig:sisd} in the $\sigma_{SD}$ vs. $M_{\tilde{\chi}_1^0}$ plane. Experimental limits from  IceCube~\cite{Aartsen:2016exj} and LZ~\cite{Akerib:2015cja} experiments. We can observe that the parameter space of this model is  accessible to these experiments. The parameter space corresponding to the $\gmu$ constraint (lower right panel), which corresponds to a stau-coannihilation and a bino like neutralino, has very low cross sections and is beyond the current search limit of these experiments.


\begin{figure}
        \includegraphics[ width=9cm, height=6.5cm]{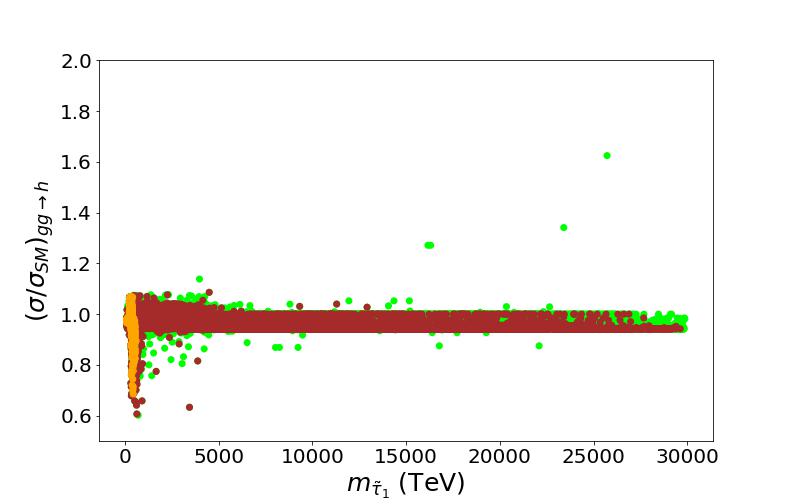}
        \includegraphics[width=9cm, height=6.5cm]{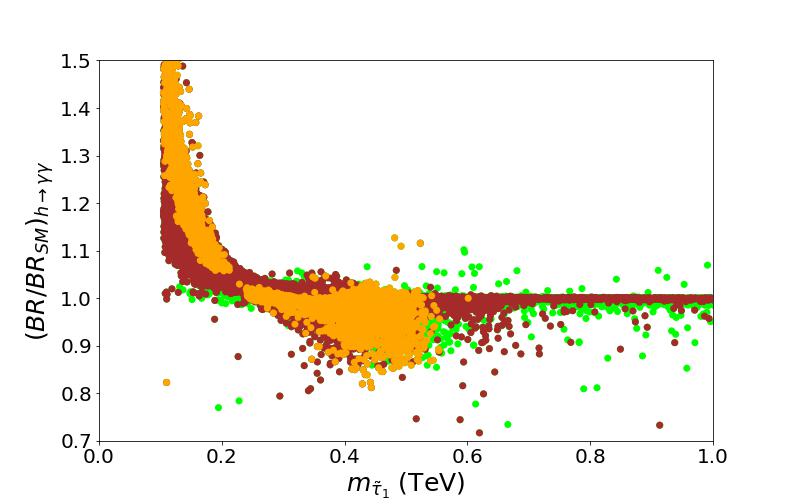} 
        \includegraphics[width=9cm, height=6.5cm]{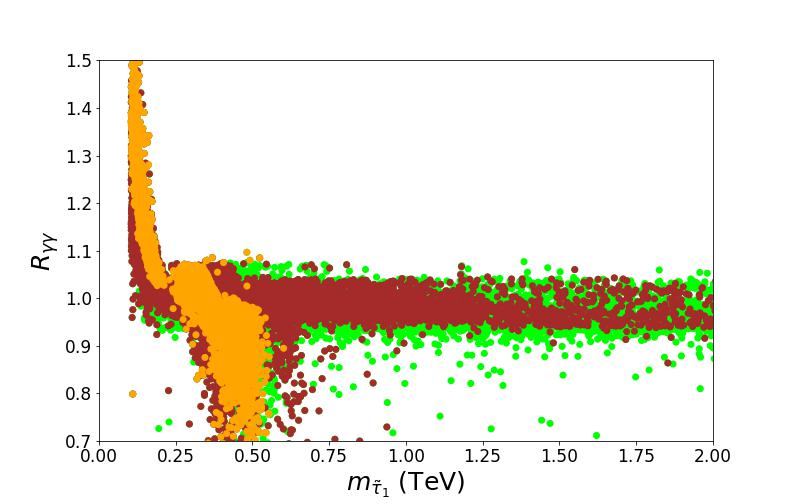}
         \includegraphics[width=9cm, height=6.5cm]{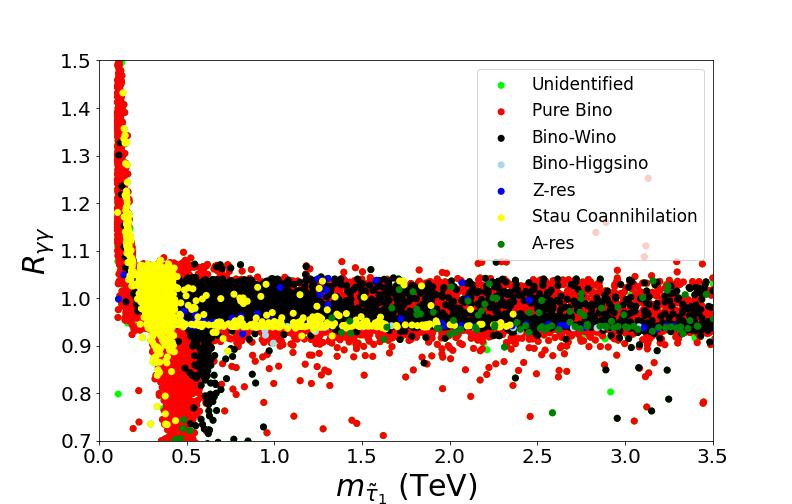}
		

\caption{Plots to display the Higgs production cross-section $(\sigma/\sigma_{SM})_{gg \rightarrow h}$, the diphoton branching ratio $(BR/BR_{SM})_{h \rightarrow \gamma \gamma}$ and $R_{\gamma \gamma}$ vs. $m_{\tilde{\tau}_1}$. Color coding for the top panels and the lower left panels is the same as in Figure \ref{fig:sparticle}. The lower right panel shows the composition of the neutralino along with various coannihilation channels shown in the plot legend.}
\label{fig:higgs-xsec}
\end{figure}

We next present the Higgs production cross section and decay in the diphoton channel. We saw in Figure \ref{fig:oh2} that the coannihilation of a light stau can lead to the desired WMAP relic abundance. In addition, the squarks are essentially decoupled in the parameter space we consider. Several studies have shown that a light stau in the decoupling limit can lead to an enhancement in the diphoton channel \cite{AdeelAjaib:2015xbe}. The ratio of the product of production cross sections times branching ratio to the diphoton final state compared to the theoretical expectation for the SM is given by \cite{AdeelAjaib:2015xbe}
\begin{eqnarray}
R_{\gamma \gamma} \equiv \frac{\sigma(h) \times Br(h\rightarrow \gamma \gamma)}{(\sigma(h) \times Br(h\rightarrow \gamma \gamma))_{SM}}.
\label{eq:ratio}
\end{eqnarray}
From Figure \ref{fig:higgs-xsec} we can see that there is essentially no enhancement in the Higgs production cross-section whereas the decay width to two photons is considerably enhanced. This leads to an enhancement in the ratio $R_{\gamma \gamma}$. We can see that a section of the parameter space that explains the g-2 constraint (\textit{orange} points) also corresponds to an enhancement in the diphoton channel. The lower right panel shows the parameter space with the composition of the neutralino and various coannihilation channels. We can see that enhancement in the diphoton channel corresponds to a light stau and a neutralino which is essentially a pure bino.

Lastly, we display three benchmark points in Table \ref{tab1}. The three points satisfy the experimental constraints presented in Section \ref{sec:parameter}. In addition, the $\gmu$ constraint is also satisfied and the relic density is consistent with WMAP. The three points have a light slepton spectrum and essentially a decoupled colored spectrum. The points exhibit stau and smuon-neutralino coannihilation to achieve the desired WMAP relic abundance. Points 2 and 3 correspond to an enhancement in the Higgs production cross-section ($R_{\gamma \gamma} > 1$) due to the light slepton spectrum\footnote{The SLHA files for these benchmark points are available on the Authors github page \url{https://github.com/Madeelajaib/nugm-nuhm2-benchmark}}.


\begin{table}[t!]\vspace{1.5cm}
\centering
\begin{tabular}{|p{3cm}|p{3cm}p{3cm}p{3cm}|}
\hline
\hline
                 	&	 Point 1 	&	 Point 2 	&	 Point 3 	\\
\hline

$m_{0} $         	&$	378.88	$&$	451.54	$&$	314.56	$\\
$M_1$         	&$	837.08	$&$	565.72	$&$	122.83	$\\
$M_2$         	&$	805.73	$&$	739.22	$&$	798.95	$\\
$M_3$         	&$	-4936.84	$&$	-4204.34	$&$	-4654.15	$\\
$A_0$         	&$	-150.77	$&$	-129	$&$	0	$\\
$\tan\beta$      	&$	42.45	$&$	40.14	$&$	14.96	$\\
$m_{H_{u}} $ 	&$	144.85	$&$	142.13	$&$	334.28	$\\
$m_{H_{d}} $ 	&$	14.97	$&$	18.98	$&$	136.21	$\\
\hline		 		 		 	
$\mu$            	&$	811	$&$	720	$&$	511	$\\

\hline		 		 		 	
$m_h$            	&$	124.54	$&$	123.93	$&$	123.55	$\\
$m_H$            	&$	1245	$&$	1990	$&$	4624	$\\
$m_A$            	&$	1237	$&$	1977	$&$	4594	$\\
$m_{H^{\pm}}$    	&$	1249	$&$	1993	$&$	4625	$\\
		 		 		 	
\hline		 		 		 	
$m_{\tilde{\chi}^0_{1,2}}$	&$	\textbf{412}, 796	$&$	\textbf{284}, 725	$&$	\textbf{88}, 786	$\\

$m_{\tilde{\chi}^0_{3,4}}$	&$	4940, 4940	$&$	4263, 4263	$&$	4628, 4628	$\\

$m_{\tilde{\chi}^{\pm}_{1,2}}$	&$	799, 4894	$&$	727, 4223	$&$	791, 4584	$\\

$m_{\tilde{g}}$  	&$	9766	$&$	8402	$&$	9229	$\\
		 		 		 	
\hline $m_{ \tilde{u}_{L,R}}$	&$	8309, 8319	$&$	7174, 7183	$&$	7849, 7859	$\\
                 		 		 		 	
$m_{\tilde{t}_{L,R}}$	&$	7229, 7351	$&$	6238, 6404	$&$	6826, 7393	$\\
                 		 		 		 	
\hline $m_{ \tilde{d}_{L,R}}$	&$	8309, 8323	$&$	7174, 7187	$&$	7850, 7865	$\\
                 		 		 		 	
$m_{\tilde{b}_{R}}$	&$	7232, 7370	$&$	6333, 6466	$&$	7351, 7777	$\\
                 		 		 		 	
\hline		 		 		 	
$m_{ \tilde{\mu}_{L,R}}$	&$	584, \textbf{455}	$&$	618, 470	$&$	587, \textbf{116}	$\\

$m_{\tilde{\tau}_{L,R}}$	&$	\textbf{457}, 957	$&$	\textbf{324}, 837	$&$	\textbf{123}, 558	$\\
                		 		 		 	
\hline		 		 		 	
$\Delta a_{\mu} \times 10^{10}$  	&$	28.20	$&$	22.55	$&$	20.06	$\\

$\sigma_{SI}({\rm pb})$	&$	1.22\times 10^{-13}	$&$	3.15\times 10^{-14}	$&$	9.45\times 10^{-15}	$\\

$\sigma_{SD}({\rm pb})$	&$	9.60\times 10^{-11}	$&$	1.78\times 10^{-10}	$&$	1.33\times 10^{-10}	$\\

$\Omega_{CDM}h^{2}$	&$	0.126	$&$	0.124	$&$	0.1	$\\

$R_{\gamma \gamma}$	&$	0.996	$&$	1.045	$&$	1.257	$\\

\hline
\hline
\end{tabular}
\caption{Masses in the table are in units of GeV. All the points satisfy the constraints presented in section \ref{sec:parameter} and have good $\gmu$ values with light sleptons. The colored sparticles are essentially decoupled for these points. As can be seen from the highlighted slepton masses, these points exhibit stau-coannihilation to achieve the desired relic density}
\label{tab1}
\end{table}


\section{Conclusion}\label{sec:conclude}

We revisited the NUGM+NUHM2 model while focusing on the long standing g-2 anomaly in the magnetic moment of the muon. We explored the parameter space of the model by highlighting the different compositions of the neutralino LSP dark matter. We find that a narrow region of the parameter space can explain the g-2 anomaly while being consistent with current experimental limits. The neutralino is predominantly a pure bino or a mixed bino-wino type in the explored parameter space. We find that a portion of the parameter space that explains the g-2 anomaly also predicts an enhancement in the Higgs production cross-section due to the light spectrum of the sleptons. 

\section{Acknowledgments}

The authors would like to thank Abdur Rehman, Bilal Riaz and Waqas Ahmed for useful discussions.

\newpage


\end{document}